\documentclass{rspublicip}
\usepackage{epsfig}
\usepackage{epsf}
\usepackage{graphicx}
\usepackage{wasysym}
\begin{document}

\newcommand{\be}{\begin{equation}}
\newcommand{\ee}{\end{equation}}
\newcommand{\bea}{\begin{eqnarray}}
\newcommand{\eea}{\end{eqnarray}}
\newcommand{\nnn}{\nonumber \\}
\newcommand{\ba}{\begin{array}}
\newcommand{\ea}{\end{array}}
\newcommand{\nuu}{\varpi}
\newcommand{\xv}{{\bf v}}
\newcommand{\xc}{{\bf c}}
\newcommand{\xD}{{\bf D}}
\newcommand{\bx}{{\bf x}}
\newcommand{\hx}{\hat{x}}
\newcommand{\hy}{\hat{y}}
\newcommand{\hz}{\hat{z}}
\newcommand{\htt}{\hat{t}}
\newcommand{\ex}{{\bf e}_x}
\newcommand{\ey}{{\bf e}_y}
\newcommand{\ez}{{\bf e}_z}
\newcommand{\et}{{\bf e}_{\theta}}
\newcommand{\er}{{\bf e}_r}
\newcommand{\xao}{\alpha_0}
\newcommand{\xo}{\mathbf{\omega}}
\newcommand{\xO}{\mathbf{\Omega}}
\newcommand{\xS}{\mathbf{\Sigma}}
\newcommand{\xep}{\langle {\bf p} \rangle}
\newcommand{\xe}{{\bf e}}
\newcommand{\xE}{{\bf E}}
\newcommand{\xI}{{\bf I}}
\newcommand{\bz}{{\bar{z}}}
\newcommand{\xA}{{\bf A}}
\newcommand{\xV}{{\bf V}}
\newcommand{\xk}{{\bf k}}
\newcommand{\xu}{{\bf u}}
\newcommand{\xq}{{\bf q}}
\newcommand{\xx}{{\wedge}}
\newcommand{\xp}{{\bf p}}
\newcommand{\xJ}{{\bf J}}
\newcommand{\xn}{{\bf n}}
\newcommand{\xj}{{\bf j}}
\newcommand{\xR}{{\bf R}}
\newcommand{\xap}{\langle{\bf p}\rangle}
\newcommand{\xw}{{\bf w}}
\newcommand{\pd}{{\partial}}
\newcommand{\pdt}{{\partial_{t}}}
\newcommand{\pdT}{{\partial_{T}}}
\newcommand{\pdz}{{\partial_{z}}}
\newcommand{\pdx}{{\partial_{x}}}
\newcommand{\pdX}{{\partial_{X}}}
\newcommand{\intab}{\int^0_{-d} dz~~}
\newcommand{\ct}{{\cal T}}
\newcommand{\cF}{{\cal F}}
\newcommand{\tM}{\tilde{M}}
\newcommand{\tN}{\tilde{N}}
\newcommand{\hM}{\hat{M}}
\newcommand{\hN}{\hat{N}}
\newcommand{\OA}{{\cal F}}
\newcommand{\tA}{\tilde{A}}
\newcommand{\tB}{\tilde{B}}
\newcommand{\tO}{\tilde{\Omega}}
\newcommand{\tP}{\tilde{\phi}}
\newcommand{\hA}{\hat{A}}
\newcommand{\hB}{\hat{B}}
\newcommand{\hO}{\hat{\Omega}}
\newcommand{\hP}{\hat{\phi}}
\newcommand{\DO}{D_{\Omega}}
\newcommand{\Sc}{S_c^{-1}}
\newcommand{\nb}{{\bf \nabla}}
\newcommand{\lapl}{{\nabla^2}}
\newcommand{\dela}{{\delta_A}}
\newcommand{\del}{{\delta}}
\newcommand{\etal}{{\it et al.}~}
\newcommand{\hf}{\frac{1}{2}}
\newcommand{\degrees}{$^{\circ}$}
\newcommand{\texto}{\rm \large}
\newcommand{\grados}{$^{\circ}C$}
\newcommand{\nl}{\hspace{0.5cm}}
\newcommand{\nota}{ \bf }
\newcommand{\CHN}{{\it Chlamydomonas nivalis}}
\newcommand{\CN}{{\it C. nivalis}}
\newcommand{\Pe}{\mbox{Pe}}
\newcommand{\nbar}{\overline{n}}
\newcommand{\prob}{P(\xR, \xp, t | \xR^\prime, \xp^\prime)}
\newcommand{\G}{{\cal G}}
\renewcommand{\r}{s}
\newcommand{\rr}{\sigma}


\title[Dispersion of microorganisms in a tube]{Dispersion of 
biased swimming microorganisms in a fluid flowing through a tube}

\author[M. A. Bees \& O. A. Croze]{Martin A. Bees and Ottavio A. Croze}
\affiliation{Department of Mathematics, University of Glasgow, Glasgow G12 8QW, U.K.
}
\label{firstpage}
\maketitle

\begin{abstract}{Taylor dispersion, gyrotaxis, algae, bacteria, spermatozoa, swimming, Poiseuille flow, biofuel, photobioreactors}
Classical Taylor-Aris dispersion theory is extended to describe the
transport of suspensions of self-propelled dipolar cells in a tubular flow.
General expressions for the mean drift and effective
diffusivity are determined exactly in terms of axial
moments, and compared with an approximation {\it a la}
Taylor. As in the Taylor-Aris case, the skewness of a finite
distribution of biased swimming cells vanishes at long
times. The general expressions can be applied to particular
models of swimming microorganisms, and thus be used to
predict swimming drift and diffusion in tubular
bioreactors, and to elucidate competing unbounded swimming drift and diffusion
descriptions.  Here, specific examples are presented for
gyrotactic swimming algae.
\end{abstract}


\section{Introduction}

\begin{figure}
\begin{center}
a)\rotatebox{0}{\includegraphics[height=5cm]{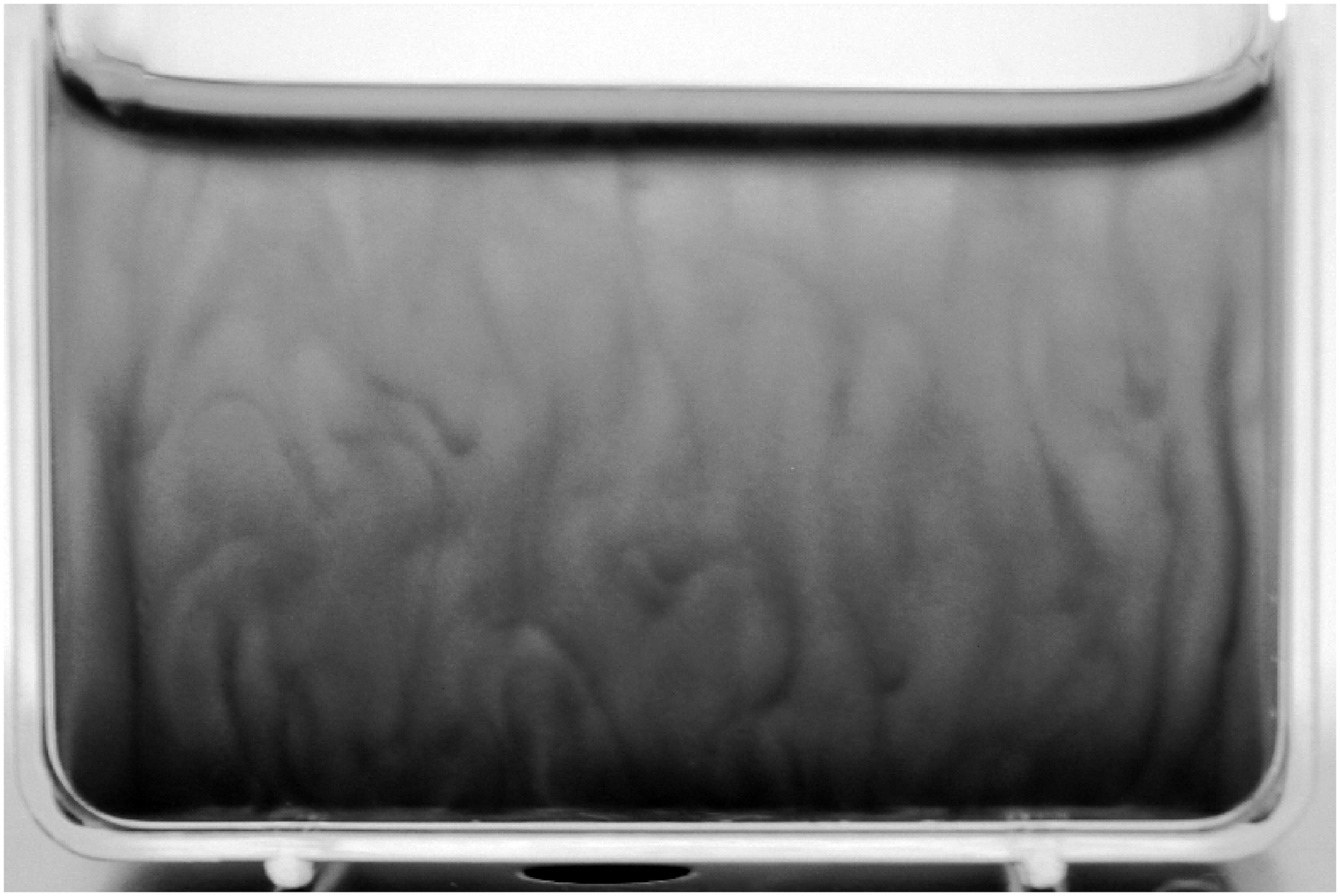}}~~
b)~\rotatebox{0}{\includegraphics[height=5cm]{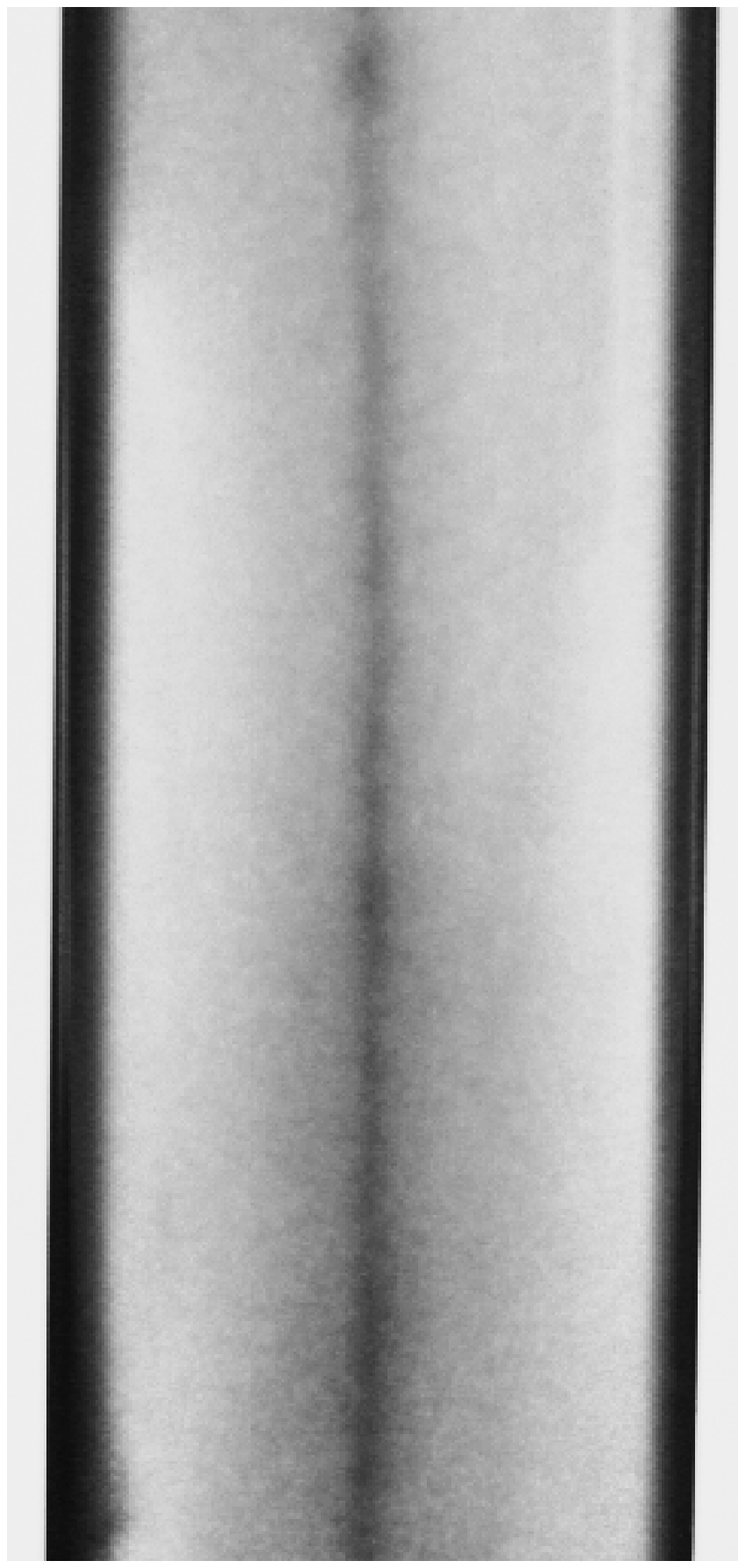}}\\
\caption{\bf Bioconvection plumes in suspensions of \CN~ in a) a culture flask (5.3 cm wide), and b) a long vertical tube of inner diameter 1.1 cm. Concentration   $\approx 1.5 \times 10^6$ cm$^{-3}$.}\label{f:biocon}
\end{center}
\end{figure}

Suspensions of swimming microorganisms, such as algae and bacteria, behave differently 
to molecular fluids.
Many microorganisms exhibit taxes, directed motion relative to 
external or local cues.  For example, various algae (e.g.~{\it Chlamydomonas} and 
{\it Dunaliella} sp.) swim
upwards on average in the dark (gravitaxis) due either to a centre-of-mass offset 
from the centre-of-buoyancy (Kessler 1986), sedimentation and anterior-posterior asymmetry in  
body/flagella (Roberts 2006), or active mechanisms (H\"{a}der \etal 2005).
This can result in aggregations of cells at upper boundaries
and, if the cells are more dense than the medium in which they swim,
overturning instabilities, termed bioconvection (Wager 1911; Platt 1961).  Furthermore, a balance between
gravitational and viscous torques can bias cells to swim towards
downwelling regions, whence their added mass amplifies the downwelling. 
This is known as a gyrotactic instability and does not require an upper boundary.
Of particular relevance here, 
Kessler (1986) observed that for a suspension of gyrotactic \CHN~ in a
vertically aligned tube, cells became sharply focused at the centre for downwelling
flow and scattered towards the edges when the flow was upwelling. 
Additionally, phototrophic algae are often phototactic 
(they swim towards weak light and
away from bright light), which can modify the instability mechanisms above,
and bacteria may exhibit chemotaxis
(e.g.~up oxygen gradients).  
In shallow containers, the above taxes
can result in very distinct bioconvection patterns, with 
characteristic lengthscales of millimeters to centimetres in just tens of seconds (Bees \& Hill 1997).  
See Pedley \& Kessler (1992) and Hill \& Pedley (2005) for reviews.
In deep cultures, one may observe long thin plumes of cells (Figure~\ref{f:biocon}) that have
a clear impact on the transmittance of light through the culture, of some relevance to 
photosynthetic algae (a ``Cheese plant effect'').

Recently there has been renewed interest in utilizing
microorganisms for fuel production.  For green algae there
are two main approaches: hydrogen production by sulphur
deficient cells (Melis \& Happe 2002) and biomass generation for biodiesel
production (Chisti 2007).  To reach economical viability, both methods
require the sustainable culture of cells, extensively and
under carefully controlled conditions. Culture systems
typically consist of arrays of tubes (vertical, horizontal
or helical) and aim to maximize light whilst maintaining
linear separation of cell stage and medium victuals. In
algal bioreactors, suspensions of algae are typically
pumped and may be bubbled or tansported turbulently to enhance nutrient/gas
mixing and reduce variance in light exposure.  These
processes, which treat a suspension of microorganisms like a
chemical fluid, are energetically
costly. Instead, efficient bioreactor designs might hope to
harness the activity of the swimming microorganisms directly in laminar flows. 
However, it is unclear how a) the mean cell
drift and b) the effective axial swimming dispersion of
cells are affected by various flow fields in the aforementioned
tube arrangements.

In a series of papers Taylor (1953, 1954a, 1954b) described
how it is possible to approximate the effective axial
diffusion of a solute in a fluid flowing through a tube.
Molecular diffusion and advection by shear each play a
distinguished role, such that the effective diffusivity is
given by $D_m + U^2 a^2/ 48 D_m$, where $D_m$ is the
molecular diffusivity, $a$ is the radius of the tube and $U$
is the mean flow speed. Subsequently, Aris (1955) formalized
the approach by solving the moment equations, extending
ubiquitously the domain of physical relevance of Taylor's
result. The methods have been extended by
many authors (e.g.~Horn \& Kipp 1971, partitioning reactions
between phases; Brenner 1980, dispersion in periodic
porous media). The value of the Taylor-Aris approach
can be measured by the wealth of practical applications 
(see Alizadeh \etal 1980). Until now, the approach 
has not been extended to suspensions of biased
swimming microorganisms in a tubular flow.  As we shall see,
it is possible to derive general expressions with few
assumptions.  However, these expressions depend upon
constitutive equations for the mean behaviour of the
cells.

We shall adopt the standard continuum approach to modelling bioconvective phenomena,
although our main result is independent of the details of these descriptions.
Recent models of dilute, gyrotactic bioconvection (Childress \etal 1975; 
Pedley \& Kessler 1990, 1992) assume that the fluid flow is governed by the Navier-Stokes equations 
with a negative buoyancy term to represent the effect of the cells on the fluid 
(Boussinesq approximation) such that
\be
\rho \frac{D{\bf u}}{Dt} = - \nabla p_{e} + n v \Delta \rho {\bf g} + \nabla \cdot \xS,
\label{eq1:m1}
\ee
where ${\bf u}(\bx,t)$ is the velocity of the suspension, $p_{e}(\bx,t)$  is the excess pressure,
$\xS(\bx,t)$ is the stress tensor, ${\bf g}$ is the acceleration due to gravity, 
$n(\bx,t)$ is the cell concentration, 
$\Delta \rho $ is the difference between the cell and fluid density, $\rho$,
and $v$ is the mean volume of a cell.  The cell Reynolds number is small (e.g. $\sim 10^{-3}$ for {\it C. nivalis}).
Furthermore, the suspension is assumed incompressible such that 
$\nabla \cdot {\bf  u} = 0  
$.
Pedley \& Kessler (1990) extended the standard Newtonian description to include
Batchelor stresses, stress associated with 
rotary particle diffusion, and swimming induced stresslets.  The first two were 
found to be qualitatively and quantitatively insignificant, and the third only plays a role in 
concentrated regions of the suspension. 
Thus in a dilute
limit one may write $\nabla \cdot \xS = \mu \nabla^2 \xu$, where $\mu$ is the fluid viscosity.  We shall employ this approximation
in explicit examples, but the main result does not require it.
Typically, over the course of a bioconvection experiment the total number of cells is conserved, 
so that one may write
\be
\frac {\partial n}{\partial t} =  - \nabla \cdot \left[ n \left ( {\bf u } + {\bf V}_c 
\right) - {\bf  D} \cdot \nabla  n \right], \label{eq1:m3}
\ee
where ${\bf V}_c ({\bf x})$ is the mean cell swimming velocity and ${\bf D}({\bf x})$ is the cell 
swimming diffusion tensor, both of which need to be determined.
At rigid boundaries, ${\cal G}$, we require a no-slip condition,
$ 
{\bf u} = {\bf 0}    \mbox{~on~}    {\cal G},
$
as well as zero cell flux normal to ${\cal G}$ (in direction ${\bf n}$), such that
$ 
{\bf n}\cdot \left( n \left( {\bf u} + {\bf V}_{c} \right) -
{\bf D}\cdot \nabla n \right) =  0 \mbox{~on~} {\cal G}.
$

To model gyrotaxis, 
Pedley \& Kessler (1987) employed a deterministic balance of gravitational and 
viscous torques on a spheroidal cell, of eccentricity $\alpha_0$, to determine the cell orientation $\xp$:
\be
\dot{\xp} = \frac{1}{2B}\left[ \xk - (\xk \cdot \xp)\xp \right] + \frac{1}{2} \xO \xx \xp + \xao \xp \cdot \xE \cdot (\xI - \xp \xp ).
\ee 
Here, $B$ is the gyrotactic reorientation time-scale of a
cell affected by external (gravitational) torques subject to
resisting viscous torques, given by $B={\mu
\alpha_{\perp}}/{2h\rho g}$, where $h$ is the
centre-of-mass offset relative to the centre-of-buoyancy
and $\alpha_{\perp}$ is the dimensionless
resistance coefficient for rotation about an axis
perpendicular to $\xp$.  $\xO$ and $\xE$ are the local
vorticity vector and rate-of-strain tensor, respectively.
These authors then wrote 
${\bf V}_c = V_s \xp$, where $V_s$ is the mean swimming speed, 
and, as for earlier models, assumed a constant
isotropic diffusion.  Pedley \& Kessler (1990) advanced this description 
by postulating that
the probability density function, $f(\xp,t)$, for orientation $\xp$ satisfies a Fokker-Planck
equation, with drift due to the various torques and a rotational diffusivity analogous to rotational Brownian motion (Frankel
\& Brenner 1991), thus taking account of biological variation of swimming stroke.  
Experimental data on cell tracking
(Hill \& H\"{a}der 1997) has provided values for the deterministic and diffusive parameters.
From $f(\xp)$, the mean swimming direction, $\xq$, is easily calculated, yielding ${\bf V}_c 
= V_s \xq$, but the cell
swimming diffusion tensor is not and requires approximation.  Pedley \& Kessler (1990)
suggested that $\xD \approx V_s^2 \tau \mbox{var}(\xp)$, where $\tau$ is a direction correlation
time, estimated from experimental data, and found asymptotic solutions for 
small flow gradients.  Bees \etal (1998) extended these solutions for all flow gradients by
expansion in spherical harmonics (employed in Bees \& Hill 1998, 1999).  However, the {\it ad hoc} nature of the diffusion 
approximation was cause for concern.  This motivated Hill \&  Bees (2002) and Manela \& Frankel (2003) to develop 
generalized Taylor dispersion theory (Frankel \& Brenner 1991), taking account of both the orientation and position
of cells swimming in a linear flow, to derive the leading order, long time, spatial diffusion tensor.
The techniques were subsequently employed by 
Bearon (2003) for dispersion of
chemotactic bacteria in a shear flow.

There are significant qualitative differences between the 
three treatments described above as vorticity is varied.  In particular, as vorticity, $\xo$, is increased 
the Fokker-Planck and the generalized Taylor dispersion approaches provide 
eigenvalues of the diffusion tensor that tend towards non-zero and zero limits, respectively.
This is due to the fundamental difference between the orientation only versus trajectory based 
descriptions.  Such qualitative differences in behaviour need to be tested with 
laboratory experiments.
One approach is to track individual microorganisms in the
very dilute limit (Hill \& H\"ader 1997; Vladimirov \etal
2004) but for a precisely prescribed shear
flow (e.g.~Durham \etal 2009).  However, such a scheme would
likely be laborious and may not easily yield significant
results for large shear rates.  A macroscopic approach would
be much preferred. In general, the coupling between cell and fluid is
bidirectional; the flow is driven by the presence of the
cells, which determines the swimming directions of the
cells.  
Controlling the flow in the manner described by Taylor may
thus be advantageous. 
There are, however, some obstacles to be
overcome.  In particular, a local distribution of cells will drive
secondary flows and lead to an effective axial diffusivity that depends on 
the axial location.
The answer is to create a flow that 
is independent of the presence of the cells.  This can be achieved by creating
a long axisymmetric plume of swimming cells and dying a small 
blob of cells within the plume (Figure 2).    
In this way, we partially uncouple the 
drift-diffusive dynamics of the dyed cells from the bulk flow-cell problem.

In the next section, we shall describe the geometry and scaling of the problem and 
introduce the method of moments.  In section~\ref{s:flow}, steady-state 
solutions of plume concentration and flow in a tube subject to a pressure gradient are
calculated.
In section~\ref{s:driftdisp}, the long-term drift and effective diffusion of a
blob of cells in a plume in a tube of circular cross-section are formulated in general terms.
The skewness of the distribution is also determined.  For general comprehension and comparison,
an argument in a vein similar to that given by Taylor (1953) is presented in 
section~\ref{s:taylor}.  The full theoretical results are then summarized in 
section~\ref{s:examples} before explicit example calculations are given.
Conclusions are presented in section~\ref{s:discussion}.

\section{Flow in a straight tube \label{s:straight}}


This analysis is applicable to the case where
the flow is independent of the axial direction.  Thus
consider the diffusion of dyed cells within a long plume.

\begin{figure}
\begin{center}
\rotatebox{0}{\includegraphics[height=7cm]{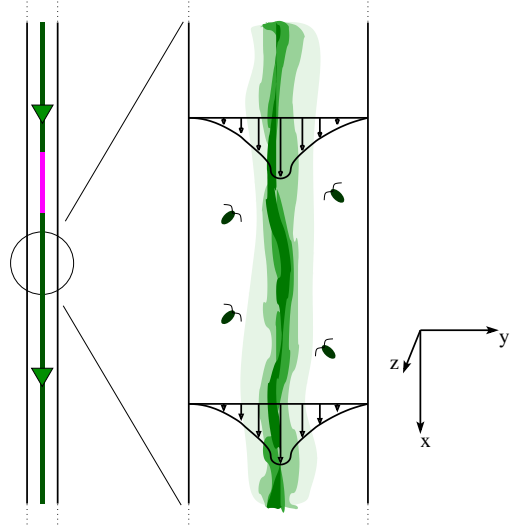}}\\
\caption{\bf Tube arrangement and coordinates.  The magenta (light grey) region on the left 
represents
dyed cells for which drift and diffusion is to be calculated.}\label{f:tube}
\end{center}
\end{figure}

We follow the notation of Aris (1955) and consider a tube with characteristic scale $a$
with axis parallel to the vertical $x$-axis (pointing in the downwards direction; see 
Figure~\ref{f:tube}).
The interior of the tube is denoted by $S$, its 
cross-sectional area by $s$ and its perimeter by $\Gamma$.  
We consider
flows, $\xu$, generated by a pressure gradient and the added mass of the algae such that 
\be
\xu(\bx_{H}) = u(\bx_{H}) \ex = U[1+\chi(\bx_{H})] \ex, \label{eq:mf}
\ee
where $U$ is the mean flow speed and $\chi$ is the flow speed relative to the mean and is assumed to be {\it only} a function 
of the cross-sectional coordinates $\bx_{H}$. Clearly, a no-slip boundary condition provides $\chi = -1$ on $\Gamma$.

Let the cell swimming diffusion tensor be of the form $D^c \xD$, where $D^c$ is its 
characteristic scale, and the mean cell swimming velocity be $V_s\xq({\bf x}_{H})$ 
(see Bees \etal 1998, where $\xq \equiv \xep$), where $V_s$ is the 
mean swimming speed.  As $\chi$ is independent of the axial direction then so are 
$\xD$ and $\xq$. This fact permits a treatment using the method of moments in a 
similar vein to that described in Aris (1955).

The cell conservation equation (\ref{eq1:m3}) can thus be written
\be\label{CC0}
\frac{1}{D^c} n_{t} = \nabla \cdot \left( \xD \cdot \nabla n \right) - \frac{U}{D^c}\left( 1 + \chi \right) n_{x}
-\frac{V_s}{D^c} \nabla \cdot \left( n \xq \right), 
\ee
where we use subscripts for partial differentiation where it is clear.
It is conducive to translate to a reference frame travelling with the mean flow,
and non-dimensionalize, such that
$\hx = (x-Ut)/a$, $\hat{\bx}_H = \bx_H/a$ and $\htt= D^c t /a^2$.
Equation (\ref{CC0}) becomes
\be
n_{t} = \nabla \cdot \left( \xD \cdot \nabla n \right) - \Pe \chi n_{x}
- \beta \nabla \cdot \left( n \xq \right), \label{CC1}
\ee
where 
\be
\Pe = \frac{Ua}{D^c}, \mbox{~~~~and~~~~}
\beta = \frac{V_s a}{D^c} \left( = \frac{a}{V_s \tau} \right),
\label{eq:Pe}
\ee 
and the hats are dropped for notational clarity.
Here, Pe is a Peclet number, which is a ratio of the rate of advection by the flow 
to the rate of swimming diffusion, 
and $\beta$ is a `swimming' Peclet number, 
a ratio of the rates of transport by swimming 
to swimming diffusion (or tube radius to swimming 
correlation length,  where $\tau$ is the direction correlation time, as typically $D^c=V_s^2 \tau$; e.g.~Pedley \& Kessler 1990).
No-flow and no-flux boundary conditions shall be applied to the solution, such that
\be 
\xu = {\bf 0} \mbox{~~~~and~~~~}
\xn \cdot \left( \xD \cdot \nabla n - \beta \xq n \right) = 0, \mbox{~~~~on~~~~} \Gamma ,
\label{nfnf}
\ee
respectively, where $\xn$ is normal to $\Gamma$. 

The $p$th moment with respect to the axial direction through $\bx_H$ 
is defined as
\be
c_p(\bx_{H},t) = \int^{+\infty}_{-\infty} x^p n(x,\bx_{H},t) dx,
\ee
provided it exists and is finite (i.e.~$x^p n(x,\bx_{H},t)\to 0$ as $x\to\pm\infty$). The 
cross-sectional average (denoted by an overbar) of this moment is written
\be
m_p(t) = \overline{c_p}=\frac{1}{s}\int\int_S c_p dS.
\ee

Henceforth, consider axisymmetric flows in a tube of circular cross-section
with radius $a$ oriented parallel to the vertical $x$-axis (pointing downwards).
Here $\chi = \chi(r)$, $\xD = \xD(r)$ and $\xq = \xq(r)$ (such that $\xq$ has no 
component in the $\et$ direction).

In cylindrical coordinates, by multiplying by $x^p$ and integrating over the 
length of an infinite pipe,
equation (\ref{CC1}) becomes
\bea
c_{p , t} & = & \frac{1}{r}\left[ r (D^{rr}c_{p,r}-\beta q^r c_p- p D^{rx}  
c_{p-1})\right]_r-  p D^{rx} c_{p-1,r} 
 \nnn
& & + p(\Pe\,\chi + \beta q^{x})c_{p-1} +p(p-1)D^{x x} c_{p-2} ,  \label{A1}
\eea
with 
\be
D^{rr} c_{p,r} - \beta q^r c_p - p D^{rx}c_{p-1} = 0 \mbox{~~~~on~~~~} r=1.\label{A1b}
\ee
Averaging over the cross-section (applying no-flux boundary conditions 
\ref{A1b}) yields
\be
m_{p , t}= -  p \overline{D^{rx} c_{p-1,r}} 
+ p\overline{ \left( \Pe \chi +\beta q^{x} \right) c_{p-1}}+p(p-1)\overline{D^{x x} c_{p-2}}. 
\label{B1}
\ee

Before deriving results for drift and diffusion in section 
\ref{s:driftdisp}, we shall solve the steady, coupled, cell conservation and 
hydrodynamic problem.

\section{Steady problem: flow and cell concentration} \label{s:flow}

Kessler (1986) demonstrated theoretically and experimentally 
that plume solutions exist in vertically aligned
tubes.  He found that the plumes are generally 
stable when a pressure gradient is applied such that the flow is downwards.  However,
varicose instabilities may arise when no pressure gradient is applied.  Here, we aim to 
avoid such instabilities and thus in the ensuing analysis 
implicitly refer to parameter regimes where plume solutions are stable.

In later sections, in order to compute the dispersion of a blob of cells 
within a plume, we require knowledge of $\chi$, the fluid velocity relative to the mean.
Hence, when $\chi(r)$ represents the steady fluid velocity induced 
by a pressure gradient and the presence of 
a swimming cell distribution that is independent of $x$,
\be
0 = \nabla \cdot \left( \xD \cdot \nabla n^* \right) 
- \beta \nabla \cdot \left( \xq n^* \right)
\ee
where $n^*$ now represents all cells in the plume, and not just 
those dyed cells for which we shall
calculate dispersion.  As $n^*_r=0=q^r$ at $r=0$, this implies that
\be
D^{rr}n^*_r = \beta q^r n^*. \label{C}
\ee
Hence, given $D^{rr}(r)$ and $q^r(r)$ we have
\be
\tilde{n} = \tilde{n}(0) \exp{\left(\beta \int^r_0\frac{q^r(\r)}{D^{rr}(\r)} d\r\right) },\label{C2}
\ee
where $\tilde{n}$ is non-dimensional cell concentration
(scaled with the average concentration, $\overline{n^*}$).
Note, for a spherical cell ($\alpha_0 = 0$), $q^r$ and $D^{rr}$ are functions of 
vorticity only, which must be in the 
$\et$ direction: $\xo =\nabla \xx \xu= -\chi_r(r) \et=\xo \et$.

In cylindrical polars the steady flow equation (\ref{eq1:m1}) in the dilute limit becomes
\be
\nabla^2 \frac{u}{U} = \frac{1}{r} \left( r \chi_r  \right)_r =   \tilde{p}_{x} - \alpha \tilde{n}, \label{eq:NS}
\ee
subject to the boundary conditions
$
\chi_r(0) = 0 
$ and
$
\chi(1) = -1. 
$
Here, the non-dimensional pressure gradient is $\tilde{p}_x = p_x a / U \mu$, and
\be
\alpha = \frac{a^2 v g \Delta \rho N}{U \nu \rho},
\ee
measures the magnitude of the effect that the cells have on the flow.  
$g$ is the acceleration due to gravity
acting in the positive $x$-direction.

Contrary to intuition, $\tilde{p}_x$ and $\alpha$ are not free parameters but 
are linked to the mean flow speed, $U$, introduced in equation~(\ref{eq:mf}).
Together they are determined by the boundary conditions on $\chi$ 
and the requirement that 
$\bar{\chi} = 0$;  the flow deviation relative to the mean is order one.
For Poiseuille flow, where $\alpha=0$, it is well known
that $\tilde{p}_x = -8$, such that $\chi = 1 - 2r^2$.

Substituting (\ref{C2}) for $\tilde{n}$, equation~(\ref{eq:NS}) can be rewritten as
\be
\frac{1}{r} \left( r \chi_r  \right)_r -   \tilde{p}_{x} = - \alpha\,\tilde{n}(0) \exp{\left( \beta \int_0^r \frac{q^r(\r)}{D^{rr}(\r)} d\r \right)}. \label{eq:al}
\ee

For spherical cells ($\alpha_0=0$), taking logs and differentiating provides
\be \label{eq:alo}
\frac{\left( \frac{1}{r} \left( r \chi_r  \right)_r \right)_r}{\frac{1}{r} \left( r 
\chi_r  \right)_r -\tilde{p}_{x}}
= \beta \frac{q^r(\omega)}{D^{rr}(\omega)} =: \gamma(\omega).
\ee
Note, differentiating removes the dependence on 
$\alpha$; to fully specify the constants of integration,
substitution back into Equation~(\ref{eq:al}) will be required.
In general, equation (\ref{eq:alo}) can be solved for $\omega$ and, thus, $\chi$ and 
$\tilde{n}$ (with application of the boundary conditions).
Later, we shall consider the simple case $\gamma(\omega) \approx A\omega$, 
for constant and negative $A$, so here we derive expressions for $\chi$ in this limit.
Equation~(\ref{eq:alo}) becomes
\be
r^2 \omega^{\prime \prime} + \left( r - A\omega r^2  \right) \omega^{\prime} 
- \left( 1 + r A \omega )   \right)  \omega =  \tilde{p}_{x} r^2 A \omega. 
\label{eq:omeq2}
\ee
$r=0$ is a singular point and so consider
$\omega = \sum^{\infty}_{m=0} b_m r^{m+Q}$,
where constant $Q$ is to be determined.
Substituting into the nonlinear equation
and examining coefficients reveals $Q=1$, for finite solutions at $r=0$.
Furthermore, the recurrence relation
\be
b_t = \frac{A\left[ \tilde{p}_x b_{t-2} + \sum^{t-2}_{m=0} b_m b_{t-m-2}(m+2) \right] }
{t(t+2)},
\ee
is forthcoming.  We require that $\omega$ is odd and, therefore,
$b_i=0$, $\forall i$ odd.
Hence, the first few coefficients are given by
\bea
b_2 = \frac{A b_0}{2^3}[\tilde{p}_x +2b_0 ], \mbox{~~~~} 
b_4 = \frac{A^2 b_0}{2^6 3}[\tilde{p}_x +2b_0 ][\tilde{p}_x +6b_0 ], \nnn
b_6 = \frac{A^3 b_0}{2^{10} 3^2}[\tilde{p}_x +2b_0 ][\tilde{p}_x +6b_0 ]
[\tilde{p}_x +8b_0 ] +
      \frac{A^3 b_0^2}{2^8 3}[\tilde{p}_x +2b_0 ]^2.
\eea
Furthermore, application of the boundary conditions yields
\be
\chi = -1 +\sum^{\infty}_{m=0} \frac{b_m}{m+2}(1-r^{m+2}).
\ee
Applying the condition $\overline{\chi}=0$ admits the result
\be 
b_0 = 4\left(1 - \sum_{m=1}^{\infty} \frac{b_m}{m+4}\right). \label{eq:b00}
\ee

Finally, substitution of $\chi$ into Equation~(\ref{eq:al}) is required to find $\alpha$
in terms of $b_m$, $m = 0,2,4,...$, and $\tilde{p}_x$.  Equation~(\ref{eq:al}) can be written
\be
\left( r \chi_r  \right)_r -   r\tilde{p}_{x} = - r\tilde{\alpha} 
\exp{\left[ -A\chi(r) \right]}, \label{eq:al2}
\ee
where $\tilde{\alpha} = \alpha \tilde{n}(0)\exp{(A\chi(0))}$ 
(evaluated with the normalization condition 
$\overline{\tilde{n}}=1$, giving
$\tilde{n}(0) e^{A\chi(0)} = 1/  {2\int_0^1 e^{-A\chi(r)} r\, dr} 
$).
Hence, substituting $\chi$ into (\ref{eq:al2}) 
and comparing coefficients at leading order in $r$ we find that
\be
\tilde{\alpha} =  \left[ 2b_0 + \tilde{p}_x\right] \exp\left\{-A\left(
1 - \sum_{m=0}^{\infty} \frac{b_m}{m+2} \right)\right\}. \label{eq:b0}
\ee
Higher orders in $r$ provide a 
check for the previously computed $b_m$, $m=2,4,6,...$.
Therefore, given $b_m$, $m=0,2,4,...$, and $\tilde{p}_x$, then $\tilde{\alpha}$ 
can be computed from (\ref{eq:b0}).

If $b_0=b_0(\tilde{\alpha})$ is required, and in the particular case that $A$ is small 
(i.e.~the cells are weakly affected by the flow; e.g.~$B$ is small), 
such that we can neglect quadratic terms in $A$ and higher, then we can
expand the transcendental equation to give
\be
b_0 = \frac{-2\tilde{p}_x + 2\tilde{\alpha}\left\{ 1+A \left( 1-\sum_{m=1}^{\infty}
\frac{b_m}{m+2} \right) \right\} }{4 + \tilde{\alpha} A}. \label{eq:ab0}
\ee

Three examples are presented below, with the profiles plotted in Figure~\ref{f:prof}.
\begin{itemize}
\item[(I)] One of the simplest cases is for $\tilde{\alpha}=0$ (i.e.~the presence of the cells does not affect the flow).  In this case, we compute 
$b_0 = -\tilde{p}_x / 2$ and $b_m=0$, $m=2,4,6,...$,
such that $\chi = -1 + \tilde{p}_x(r^2-1)/4$, which is Poiseuille flow.
Equation~(\ref{eq:b00}) gives $b_0=4=-\tilde{p}_x/2$, as would be expected. 
\item[(II)] With $\tilde{\alpha}\neq 0$ and $A<0$ small (i.e. a broad plume), but a zero pressure gradient, 
$\tilde{p}_x=0$, then $b_2 = Ab_0^2/4$, $b_m = O(A^2)$, $m=4,6,8,...$.  
and equations~(\ref{eq:b00}) and (\ref{eq:ab0}) provide
$b_0 = 6(-1 \pm \sqrt{1+8A/3})/2A + O(A^2)
= \tilde{\alpha} \left[ 1 + \left(1-\tilde{\alpha}/4 \right) A \right] + O(A^2)$.
Thus $\chi = -1 + b_0(1-r^2)/2 + Ab_0^2(1-r^4)/16 + O(A^2)$.  
Two solutions are possible: a simple positive flow (mode 1), and one with upwelling towards the edge of the tube (mode 2).
For zero pressure gradient, a closed form, mode 1 solution is known.
Kessler (1986) noted that
$\tilde{n}=\tilde{n}(0)/(1+C_1\tilde{n}(0)r^2)^2$ is a solution, for constant $C_1$.
Applying the condition $\overline{\tilde{n}}=1$ gives $C_1=\tilde{n}(0)-1$.  Substituting
this solution back into the governing equation reveals that $\tilde{n}(0)$ is determined by
the constrained parameter $\tilde{\alpha}$, as should be the case, 
in the same way that the mean velocity is linked to 
the pressure gradient in Poiseuille flow.
This closed-form profile is approached by the above mode 1 profile
with truncated sums (not shown). 
\item[(III)] The case $A<0$ (small; a broad plume) and $\tilde{p}_x \neq 0$ is also of interest, and has not 
previously been investigated.
If $A=-\frac14$ and $\tilde{p}_x=-6$, then we calculate 
$b_2=-b_0(b_0-3)/2^4$, $b_4 = b_0(b_0-3)(b_0-1)/2^8$, and
$b_6=-b_0(b_0-3)(b_0-1)(4b_0-3)/(2^{13}3) - b_0^2(b_0-3)^2/(2^{12}3)$. From
equation~(\ref{eq:b00}) we compute the two solutions $b_0 \approx 4.179$ and 
$21.931$. Again, the mode 2 solution corresponds to a flow with 
upwelling near the edge of the tube.
The corresponding $\tilde{\alpha}$ can be evaluated from equation~(\ref{eq:b0}).  
Hence, for the mode 1 solution, $\chi = -1 + 4.179(1-r^2)/2 - 0.308(1-r^4)/4 + 0.0612(1-r^6)/6 - 0.0107(1-r^8)/8 + ...$.
\end{itemize}
\begin{figure}
\begin{center}
\rotatebox{0}{\includegraphics[width=8cm]{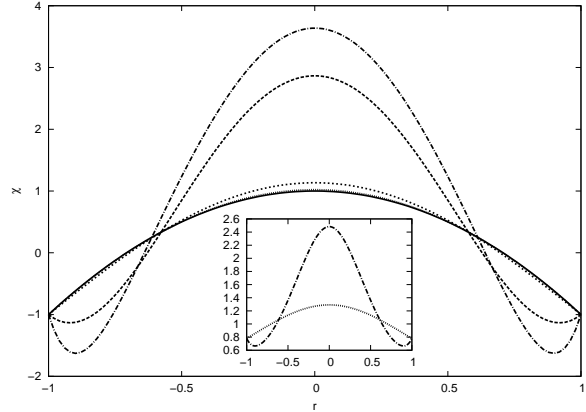}}\\
\caption{\bf Flow profiles for three cases (see text; extended to cover the 
width of the tube): 
$\tilde{\alpha}=0$, $\tilde{p}_x = -8$ (solid); 
$\tilde{\alpha}\neq 0$, $\tilde{p}_x = 0$, $A=-\frac14$ (two dashed curves);
$\tilde{\alpha}\neq 0$, $\tilde{p}_x = -6$, $A=-\frac14$ (dotted and dot-dashed curves).  
Selected broad cell distributions, $\tilde{n}/ \tilde{n}(0)e^{\chi(0)}$, are also plotted (inset; see text).}\label{f:prof}
\end{center}
\end{figure}

\section{Dispersion in a tube of circular cross-section} \label{s:driftdisp}

In this section, we place no restrictions on the cell shape (e.g.~spheroidal cells) and form of $\chi(r)$, 
${\bf q}(r)$ and ${\bf D}(r)$, and find general expressions for the drift and 
effective diffusion of a blob of dyed cells within an existing plume.

\subsection{Cell conservation and drift}

For $p=0$ equation (\ref{B1}) gives $m_{0,t} = 0$, so that $m_0$ is a constant (i.e. number of cells is conserved).  We fix $m_0=1$ (and remember that these cells represent dyed cells diffusing within a plume of other cells).  Equation (\ref{A1}) with $p=0$  implies that 
\be
c_{0 , t}= \frac{1}{r}\left[ r (D^{rr}c_{0,r}-\beta q^r c_0 )\right]_r,\label{A10}
\ee
with boundary condition
$
D^{rr} c_{0,r}-\beta q^rc_0=0$ on $r=1 \label{A10b}
$.
The solution takes the form
\be
c_0 = R_0^0(r) + \sum^{\infty}_{n=1} R_{n}^0(r)T_{n}^0(t),
\ee
where $T_{n}^0 = \exp{\left( -\gamma_{n}^2 t\right)}$ and $R_{n}^0$ satisfies
\be
r D^{rr}R_{n}^{0\prime \prime} +  \left( D^{rr} +r D^{rr\prime}  \right) R_{n}^{0\prime}
+  \left( -\beta q^r - r\beta q^{r\prime} + \gamma_{n}^2 r  \right) R_{n}^0 = 0, \label{G}
\ee
subject to the initial conditions.
The solution for $R_0^0$, such that $\overline{R_0^0}( =\overline{c_0})=1$, is
\be\label{R00}
R_0^0(r) = \exp{\left( \beta \int^r_0 
\frac{q^r(\r)}{D^{rr}(\r)}d\r \right)}\left\{\overline{ \exp{\left( \beta \int^r_0 \frac{q^r(\r)}{D^{rr}(\r)}d\r \right)}}\right\}^{-1}.
\ee
Note also that $\overline{R_n^0} = 0$, $n \neq 0$. Putting $p=1$ in (\ref{B1}) gives
\be
m_{1,t} = \Lambda_0 + \sum_{n=1}^{\infty} \exp{\left( -\gamma_n^2 t\right) } \Lambda_n,
\mbox{~~~where~~~}
\label{LamdanDef}
\Lambda_n = -\overline{D^{rx}R_{n}^{0\prime}}
+ \overline{\left( \Pe \chi + \beta q^{x} \right) R_{n}^0}.
\ee
In particular, it is clear that
\be
\lim_{t \rightarrow \infty} m_{1,t} = \Lambda_0 = -\overline{D^{rx}R_{0}^{0\prime}}
+ \overline{\left( \Pe \chi + \beta q^{x} \right) R_{0}^0}. \label{H} 
\ee
This means that the mean of the blob of dyed cells will move at a speed of $\Lambda_0$ 
relative to the mean flow.  Hence,
\be
m_1(t) =  \Lambda_0 t+ \sum_{n=1}^{\infty} \frac{1}{\gamma_{n}^2}\left(  1 - 
\exp{\left(-\gamma_n^2 t \right)} \right) \Lambda_{n} , \label{I}
\ee
where we have used $m_{10} = 0$.  The first term of
$\Lambda_0$ in (\ref{H}) is associated with a diffusive flux, the second
with advection of the cells heterogeneously distributed near
the axis of the tube, and the third to swimming in the
vertical direction relative to the fluid motion. At long times we expect
\be
m_{1\infty}(t) =   \Lambda_0 t+ \sum_{n=1}^{\infty} \frac{\Lambda_{n}}{\gamma_{n}^2}   , 
\label{driftinf}
\ee

\subsection{Effective diffusion}

With $p=1$, equation (\ref{A1}) implies that 
\be \label{A20}
c_{1 , t}-\frac{1}{r}\left[ r (D^{rr}c_{1,r}-\beta q^r c_1- D^{rx}  c_{0} )\right]_r= -  
D^{rx} c_{0,r} + (\Pe\,\chi + \beta q^{x})c_{0},
\ee
with boundary condition
\be
D^{rr} c_{1,r}-\beta q^rc_1-D^{rx} c_0=0 \mbox{~~~~on~~~~} r=1. \label{A20b}
\ee
The solution of this equation can be constructed in {\bf three parts}.
\begin{enumerate}
\item{Particular integral from $R_0^0(r)$ in $c_0$.  It satisfies
\be
\hspace{-1mm}-c_{1,t}^1 + \frac{1}{r}\left[  r \left( D^{rr}  c^1_{1,r} 
- \beta  q^{r}  c^1_{1} -D^{rx} R_0^0 \right) \right]_r 
= D^{rx} {R_0^0}^{\prime} - \left( \Pe \chi + \beta q^{x} \right) R_0^0. \label{J}
\ee}
\item{Particular integral from the rest of the terms 
$ R_n^0(r)\exp{(-\gamma_n^2 t)} $, $n\neq0$, in $c_0$.
\bea
-c_{1,t}^2 + \frac{1}{r}\left[  r \left( D^{rr}  c^2_{1,r} 
- \beta  q^{r}  c^2_{1} -D^{rx} R_n^0(r)e^{-\gamma_n^2 t} \right) \right]_r & & \nnn
= D^{rx} {R_n^0(r)}^{\prime}e^{-\gamma_n^2 t}-\left( \Pe \chi + \beta q^{x} \right) R_n^0(r)e^{-\gamma_n^2 t}. & & \label{J2}
\eea 
It is quite clear that solutions to (\ref{J2}) are of the form 
$S_n(r) \exp{(-\gamma_n^2 t)} $, where $S_n$ satisfy no-flux boundary conditions and are found by solving
\bea
\gamma_n^2 S_n(r) + \frac{1}{r}\left[  r \left( D^{rr}  S_n(r)^{\prime} 
- \beta  q^{r}  S(r)-D^{rx} R_n^0(r) \right) \right]_r & & \nnn 
= D^{rx} {R_n^0(r)}^{\prime}-\left( \Pe \chi + \beta q^{x} \right) R_n^0(r). 
& & \label{J2S}
\eea 
As we are interested in long-time behaviour, we do not solve for $S_n(r)$, but later will require its cross-sectional average. This can be found by averaging both sides of (\ref{J2S}) and using the boundary conditions (\ref{A20b}), to give
\be\label{Snavg}
\overline{S_n}=-\frac{1}{\gamma_n^2}\left[-\overline{D^{rx} {R_n^0}^{\prime}}+\overline{\left( \Pe \chi + \beta q^{x} \right) R_n^0}\right]=-\frac{\Lambda_n}{\gamma_n^2}.
\ee
}
\item{Complementary function.  
Solutions of (\ref{A20}) without terms in $c_0$ that satisfy (\ref{A20b}) are of the form
$
A^1_n R_n^0(r)e^{-\gamma_n^2 t},
$
where $A_n^1$ are constants.}
\end{enumerate}

For item 1, to calculate $c_1^1$, we rewrite the equation
as 
\be
c_{1,t}^1 r - \left[  r \left( D^{rr}  c^1_{1,r} 
- \beta  q^{r}  c^1_{1} -D^{rx} R_0^0 \right) \right]_r 
= r\left(-D^{rx} {R_0^0}^{\prime} 
+ \left( \Pe \chi + \beta q^x \right) R^0_0  \right) =: \lambda_0(r), \label{PI}
\ee
Recalling that
$R_0^0$ satisfies $D^{rr} {R_0^0}^{\prime} - \beta q^r R_0^0=0$,
let $c^1_1(r,t) = \left[ Mt +f(r) \right] R^0_0$, where $M$ is a constant
and $f(r)$ is a function of $r$.
Then, (\ref{PI}) becomes 
\be
\left[ r(f^{\prime} D^{rr} R_0^0 - D^{rx}R^0_0)  \right]^{\prime} 
= -\lambda_0 + M R^0_0 r,
\ee
an equation independent of $t$. Hence, integrating once provides
\be\label{feq}
r(f^{\prime} D^{rr} R_0^0 - D^{rx}R^0_0)=  -\frac12 \Lambda_0^*(r) + \frac12 M m^*_0(r), 
\ee
where
\be
\Lambda_0^*(r) = 2 \int^r_0 \lambda_0(\r) d\r= 2 \int^r_0  \r\left( -D^{rx} {R_0^0}^{\prime} 
+ \left( \Pe \chi + \beta q^x \right) R^0_0  \right)d\r,\label{lambdastar}
\ee
\be
m_0^*(r) = 2 \int^r_0 \r R^0_0(\r) d\r,\label{m0star}
\ee
$\Lambda_0^*(1) =\Lambda_0$ and $m_0^*(1) = 1$. Applying the no-flux boundary condition (\ref{A20b}) to (\ref{feq}) yields
$M = \Lambda_0$. Integrating (\ref{feq}) again provides
\be
c_1^1(r,t) = R_0^0(r) \left( \Lambda_0 t + f(r)\right) 
= \left[ \Lambda_0 t + J(r) - \Phi(r) \right] R^0_0(r),
\ee
where
\be
J(r)=\int^r_0 \frac{D^{rx}(\r)}{D^{rr}(\r)}d\r \mbox{~~~~and~~~~} 
\Phi(r) = \frac12 \int^r_0 \left( \frac{\Lambda_0^*(\r) 
- \Lambda_0 m_0^*(\r)}{\r D^{rr}(\r)R^0_0(\r)}  \right) d\r. \label{Phidef}
\ee
Hence, the complete solution $c_1 = c^1_1 +c^2_1 +c^3_1$ is given by 
\bea\label{c1solution}
c_1 =  \left[ \Lambda_0 t +J(r) - \Phi(r) \right]R^0_0 
+ \sum_{n=1}^{\infty} S_n(r) e^{-\gamma_n^2 t}
+ \sum_{n=0}^{\infty} A_n^1 R^0_n(r) e^{-\gamma_n^2 t}. 
\eea
$A_n^1$ are chosen to fit the initial data $c_{10}(r)$.  
In particular, the value of $A^1_0$ is fixed by the initial condition 
$m_{10}=\overline{c_{10}}=0$. With $\overline{R^0_n}=0$, $n\neq 0$, this implies
\be\label{A10def}
A^1_0 = \left[ \OA 
- \sum_{n=1}^{\infty} \overline{ S_n(r) } \right], \mbox{~~~~where~~~~}
\OA = \overline{ [\Phi(r)-J(r)] R^0_0}.
\ee
Thus the axial mean eventually is distributed across the tube as
\bea\label{c1inf}
c_{1\infty}(r,t) &=& \left[ \Lambda_0 t - \sum_{n=1}^{\infty} \overline{ S_n(r) }\right]R^0_0(r)+\left[ 
J(r) - \Phi(r) +  \OA \right]R^0_0(r) 
\nnn &=& 
\left( A_0^1+\Lambda_0t +f(r) \right)R_0^0(r).
\eea
After averaging across the cross-section we obtain
$\overline{c_{1\infty}}=m_{1\infty} = \Lambda_0 t- \sum_{n=1}^{\infty} \overline{ S_n(r) }$,
which can be compared with the earlier equation for the long-time limit of $m_1$ 
(\ref{driftinf}), and allows the identification
$\overline{ S_n(r) }=-\Lambda_n/\gamma_n^2$ consistent with equation (\ref{Snavg}).

Putting $p=2$, substituting the long-time solutions for $c_1$ and $c_0$ in (\ref{B1}),
and using definition (\ref{LamdanDef}) for $\Lambda_0$, gives
\bea
m_{2,t} &=& 
  - 2\overline{D^{rx} \left[ (J-\Phi+\OA) R^0_0 \right]^{\prime}} 
+ 2\overline{ \left( \Pe \chi +\beta q^{x} \right) (J-\Phi+\OA) R^0_0 }
\nnn & & 
+2 \Lambda_0 \left[ \Lambda_0 t - \sum_{n=1}^{\infty} \overline{ S_n(r) }\right] 
+2\overline{D^{xx} R_0^0} + \mbox{O}\left\{ e^{-\gamma^2 t} \right\}. \label{eq:m2}
\eea 

If $D_{e}$ is the effective axial diffusion then one may define $
D_{e} = \lim_{t\rightarrow \infty} \frac12 \frac{dV}{dt}$,
where $V$ is the variance
($V = \frac{1}{s}\int^{\infty}_{-\infty} \int_S (x-\overline{x})^2 n ds dx$). Then 
\bea \label{De} 
D_{e}&=& \lim_{t\rightarrow \infty} \frac12 \frac{d}{dt}(m_{2} - m^2_{1})\\
& =& - \overline{D^{rx} \left[ (J-\Phi+\OA) R^0_0 \right]^{\prime}} + \overline{ \left( \Pe 
\chi +\beta q^{x} \right) (J-\Phi+\OA) R^0_0}+ \overline{D^{xx} R_0^0} \nonumber
\eea

\subsection{Third moment and approach to normality}

With $p=3$ equation (\ref{B1}) becomes
\be
m_{3 , t}= -  3 \overline{D^{rx} c_{2,r}} 
+ 3\overline{ \left( \Pe \chi +\beta q^{x} \right) c_{2}}+6\overline{D^{x x} c_{1}},
\label{m3eq}
\ee
where  $c_{2}$ is a solution to (\ref{A1}) with $p=2$,
\bea\label{p2eq}
c_{2 , t} & = & \frac{1}{r}\left[ r (D^{rr}c_{2,r}-\beta q^r c_2- 2 D^{rx}  c_{1})\right]_r-  2 
D^{rx} c_{1,r}
 \label{c2eq} \\ \nonumber 
& & + 2(\Pe\,\chi + \beta q^{x})c_{1} +2D^{x x} c_{0} , 
\eea
subject to
$D^{rr} c_{2,r} - \beta q^r c_2 -2 D^{rx}c_{1} = 0$ on 
$r=1$.
The long time solution to equation (\ref{p2eq}) has the form
\be\label{c2infty}
c_{2\infty}(r,t) = \left[2 D_e t + \Lambda_0^2 t^2 +2 A_0^1 \Lambda_0 t + 
B^1_0\right]R_0^0(r)+2(\Lambda_0 t + A_0^1)f(r) R_0^0(r)+g(r)R^0_0(r),
\ee
where $B^1_0$ is a constant determined by the initial distribution of dyed cells and function $g(r)$ can be established 
after some algebra.
Substituting (\ref{c1inf}) and (\ref{c2infty}) into (\ref{m3eq}) gives
\be\label{m3eq2}
\frac{m_{3 , t}}{3}= \Lambda_0 \left[2 D_e t + \Lambda_0^2 t^2 +2 A_0^1 \Lambda_0 t + 
B^1_0\right]  + 2(\Lambda_0 t + A_0^1)(D_e - \OA \Lambda_0) +\overline{H(r)}
\ee
where $H(r)=-D^{rx}  \left(g R^0_0\right)^{\prime} +\left( \Pe \chi +\beta q^{x} \right) g 
R_0^0+2D^{x x} f R_0^0$ and we have used the definitions (\ref{H}) and (\ref{De}) for 
$\Lambda_0$ and $D_e$  respectively. 
Rearrangement 
and integration thus provides
\be\label{m3eq5}
m_{3 }-3m_{1}m_{2} +2 m_{1}^3=  3\left[\overline{H(r)} - 2D_e \overline{f R_0^0}- \Lambda_0 
\overline{g(r)R^0_0}\right]\,t+ \mbox{const},
\ee
yielding the absolute skewness, $\sqrt{\zeta}$, of the concentration distribution, with
\be\label{m3eq6}
\zeta(t)=\frac{(m_{3 }-3m_{1}m_{2} +2 m_{1}^3)^2}{(m_{2\infty}- m_{1\infty}^2)^3}
=\frac{9\left[\overline{H(r)} - 2D_e \overline{f R_0^0}- 
\Lambda_0 \overline{g(r)R^0_0}\right]^2}{8D_e^3} \frac{1}{t} + O(\frac{1}{t^2}).
\ee
Hence, the skewness of the distribution 
decays to zero as $t^{-1/2}$ as in classical Taylor-Aris dispersion; 
at long times we expect a 
Gaussian profile for the algal blob averaged across the cross-section. 

\section{Mean drift and effective diffusion `a la Taylor'} \label{alaTaylor}\label{s:taylor}

It is instructive to re-derive approximations to equations (\ref{H}) and (\ref{De})  
from (\ref{CC1}) using an approach similar to that of Taylor (1953). 
We use Taylor's approximations without a rigorous attempt to defend them.
We begin by assuming that the cell concentration can be written as a superposition of  
the cross-sectionally averaged concentration, 
$\overline{n}=\overline{n}(x,t)(\equiv\frac{1}{s}\int\int_S n dS$), given that it 
is well-defined,
and a term $\delta n=\delta n(x,r,t)$ for the radial variation, such that
\be\label{Taysep}
n(x,r,t)=\overline{n}(x,t)+\delta n(x,r,t).
\ee
Substituting (\ref{Taysep}) into (\ref{CC1}) we find
\bea\label{Taysep2}
\nbar_{t}+\delta n_t&=& \frac{1}{r}\left\{ r [D^{rr} \delta n_r -\beta q^r 
(\nbar+\delta n)+D^{rx} (\nbar_{x}+ \delta n_x) ]\right\}_r \nnn & & 
 + D^{rx} \delta n_{rx} - (\Pe \chi + \beta q^{x}) (\nbar_{x}+\delta n_x) + D^{x x} 
(\nbar_{x x}+\delta n_{x x})
\eea
subject to 
$D^{rr} \delta n_r-\beta q^r (\nbar+\delta n)+D^{rx} (\nbar_{x}+ \delta n_x) = 0$ on $r=1$. 
Then, taking the cross-sectional average of both sides of (\ref{Taysep2}) gives
\bea\label{Taysep3}
\nbar_{t}&=& \overline{D^{rx} \delta n_{rx}} -  \beta \overline{q^{x}}\,\nbar_{x}- 
\overline{(\Pe \chi + \beta q^{x}) \delta n_x} +\overline{D^{x x}}\,\nbar_{x x}+\overline{D^{x 
x} \delta n_{x x}},
\eea
where we have used $\overline{\delta n}=0=\overline{\chi}$ and the boundary condition. 
The aim is to express $\delta n$ as a function of $\nbar$
to write (\ref{Taysep3}) in the form of an advection-diffusion equation for 
$\nbar$.
First, subtract (\ref{Taysep3}) from (\ref{Taysep2}) to obtain
\bea\label{Taysep4}
\delta n_t&=& \frac{1}{r}\left\{ r [D^{rr} \delta n_r -\beta q^r (\nbar+\delta n) +D^{rx} 
(\nbar_{x}+ \delta n_x)]\right\}_r +\beta \overline{q^{x}}\,\nbar_{x} \\ & &
- (\Pe \chi + \beta q^{x}) \nbar_{x}
- [ (\Pe \chi + \beta q^{x}) \delta n_{x}-\overline{(\Pe \chi + \beta q^{x}) 
\delta n_x}]+ \nnn & & 
 +D^{rx} \delta n_{rx}-\overline{D^{rx} \delta n_{rx}}  
+(D^{x x}-\overline{D^{x x}})\nbar_{x x} +D^{x x} \delta n_{x x} 
-\overline{D^{x x} \delta n_{x x}}. \nonumber
\eea
Next, with Taylor (1953, 1954b), we make the following assumptions: (1) axial contributions to diffusion are 
negligible with respect to the radial ones and axial advection ($\nabla^2_x n\ll \nabla^2_r n$ 
and $\nbar_x$); (2) concentration gradients in the axial direction are independent of radial 
position ($\delta n_x\approx 0$; $n_x\approx \nbar_x$); (3) transients decay rapidly; and (4), for simplicity,  
radial concentration fluctuations about the mean are small, $\delta n \ll \overline{n}$. 
Note that the
last assumption is not necessary, and is made only for illustrative convenience.
Equation (\ref{Taysep4}) then reduces to
\be\label{Taysep5}
\frac{1}{r}\left\{ r [D^{rr} \delta n_r -\beta q^r \nbar +D^{rx} \nbar_{x} ]\right\}_r=  [ (\Pe 
\chi + \beta q^{x})  - \beta \overline{q^{x}}\,]\nbar_{x} 
\ee
subject to $D^{rr} \delta n_r -\beta q^r \nbar+D^{rx} \nbar_{x}= 0$,
on $r=1$.
Equation (\ref{Taysep5}) thus gives
\be\label{Taysep6}
\delta n=\delta R^0_0\,\nbar-(J-\phi+\alpha)\nbar_x,
\ee
where $\delta R^0_0= \beta \int^r_0 \frac{q^r}{D^{rr}}dr$, and
$J=\int^r_0 \frac{D^{rx}}{D^{rr}}dr$, as in equation (\ref{Phidef}).  Furthermore,
\be
\phi(r) =\frac{1}{2}\int^r_0  \frac{ 2\int^{\r}_0 \rr (\Pe \chi(\rr)
+ \beta q^{x}(\rr) - \beta\overline{q^x})d\rr}{\r D^{rr}(\r)}  d\r,
\ee
and $\alpha=\overline{(\phi-J)}$ is a constant obtained by imposing $\overline{\delta n}=0$.
Using (\ref{Taysep6}), (\ref{Taysep3}) reads  
\be\label{Taysep7}
\nbar_{t}+ \Lambda_0\,\nbar_{x}= D_e\,\nbar_{x x},
\ee
where we neglect terms of order $\nbar_{xxx}$, 
consistent with previous approximations, and 
\begin{eqnarray}
\Lambda_0 &=&  - \overline{D^{rx} \delta R_{0}^{0\prime}}+ \overline{\left( \Pe \chi + \beta 
q^{x} \right) \delta R_{0}^0}+\beta \overline{q^{x}},\\
D_{e} &=&  - \overline{D^{rx}(J-\phi+\alpha)^{\prime}} +\overline{ (\Pe \chi +\beta q^{x}) 
(J-\phi+\alpha)}+\overline{D^{xx}}.
\end{eqnarray}
The above equations are limiting forms of (\ref{H}) and (\ref{De}). To see this, expand 
$R_{0}^0\simeq (1+\delta R_0^0)$, where $\delta R_0^0=\beta \int^r_0 
\frac{q^r(\r)}{D^{rr}(\r)}d\r\ll 1$ (implying a broad distribution across the tube). Substituting 
into (\ref{H}) and (\ref{De}) and neglecting terms of order $(\delta R_0^0)^2$, leads to the 
above expressions. 
As earlier, there is a drift of cells relative
to the flow due to swimming, diffusion and cell weighted average of the flow. 


\section{Examples of dispersion} \label{s:examples}

\subsection{Summary of drift and effective diffusion}\label{s:summary}

To recap our main results, the drift, $\Lambda_0$, and effective axial diffusivity, $D_e$, 
of a dyed blob of algae within an 
axisymmetric algal plume in a tube of circular cross-section are given by
\bea
\Lambda_0 &=&  -\overline{D^{rx}R_{0}^{0\prime}}
+ \overline{\left( \Pe \chi + \beta q^{x} \right) R_{0}^0}, \label{drifteff} \\
D_{e}&=& - \overline{D^{rx} \left[ (J-\Phi) R^0_0 \right]^{\prime}} 
+ \overline{ \left( \Pe \chi +\beta q^{x} -\Lambda_0\right) (J-\Phi) R^0_0}+ \overline{D^{xx} R_0^0}, \label{Deff}
\eea
where Pe and $\beta$ are Peclet numbers (equation~\ref{eq:Pe}).

To evaluate the above expressions we require the flow field relative to the mean, 
$\chi(r)$, and constitutive 
equations for the mean cell swimming direction, $\xq(r)$, and swimming
diffusion tensor, $\xD(r)$.  Expressions for $\chi(r)$ are obtained in 
section~\ref{s:flow}, 
and $\xq(r), \xD(r)$ are available from solutions to deterministic or 
statistical models of gyrotaxis (Pedley \& Kessler 1987, 1990; Bees \etal 1998; Hill 
\& Bees 2002; Manela \& Frankel 2003).

The base distribution of cells, $R_0^0(r)$, is defined by equation (\ref{R00}).
Furthermore, the functions $J(r)$ and $\Phi(r)$ are computed from equations~(\ref{Phidef}),
and require the functions $\Lambda_0^*(r)$ and $m_0^*(r)$ defined by equations 
(\ref{lambdastar}) and (\ref{m0star}), respectively.

\subsection{The limit to classical Taylor-Aris dispersion}

A useful check on the results is to reduce them to the original `non-swimming' form of Taylor 
(1953) and Aris (1955). The original molecular solutes were assumed to diffuse isotropically 
(with no biased motion) and have no influence on the flow. Hence, put 
$D^{xx}=1=D^{rr}$, $D^{rx}=0$ (thus $J(r)=0$) and $q_i=0$.  For a circular pipe, Poiseuille flow provides $\chi(r) = 
1-2r^2$. Thus, $\Phi(r) =\Pe\frac{1}{2}\int^r_0\frac{1}{\r}  \left(\int^{\r}_0 \rr \chi(\rr) d\rr \right) 
d\r=\Pe\left(\frac{r^2}{4}-\frac{r^4}{8}\right)$, so that $\overline{\chi \Phi}=\Pe/48$ and  
$R^0_0 = 1$.  Then the 
effective transport coefficients (\ref{drifteff}) and (\ref{Deff}) reduce to
\be\label{MomentsVDTay}
\Lambda_0 =  0\mbox{~~~~and~~~~}
D_{e} = 1 + \Pe^2 \frac{1}{48},
\ee
the classical Taylor-Aris result. In this same limit, equation (\ref{c1inf}) for the 
the centre of mass of the solute distribution at long times reduces to $c_{1\infty} = - \Phi(r) 
+ \overline{ \Phi}+ \sum_{n=1}^{\infty}(\Lambda_n/\gamma_n^2)$ so that 
$m_{1\infty}=\overline{c_{1\infty}}=\sum_{n=1}^{\infty}(\Lambda_n/\gamma_n^2)$, where 
$\overline{ \Phi}=\Pe/12$ and $\Lambda_n=\Pe\overline{\chi R_n^0}$. Thus $c_{1\infty} = 
m_{1\infty}+\Pe\left({1}/{12}-{r^2}/{4}+{r^4}/{8}\right)$,
consistent with Taylor-Aris.

\subsection{Poiseuille flow limit for weak ($\eta\omega \ll 1$) and 
strong gyrotaxis ($\eta\omega \gg 1$)}

As a second example, consider a simple Poiseuille flow not affected by the presence of the
cells. Then, $\chi = 1-2r^2$ and $\omega =-\chi_r= 4r$.  We consider the limits of weak and 
strong gyrotaxis quantified by the ratio of the timescale for reorientation by the flow,  
$\Omega^{-1}=(U/a)^{-1}$, and the characteristic time-scale for reorientation of a cell by 
gravity against viscous resistance,  $B=\frac{\mu v a_\perp}{2 m g h}$, the gyrotactic 
reorientation time. The ratio $\eta=B\Omega$ is called the non-dimensional
gyrotaxis parameter. Here, we consider two limits for which analytic solutions are known 
for the Fokker-Planck equation governing the probability distribution for the  
cell orientation $\xp$ of spherical cells, due to Pedley \& Kessler (1992) and Bees \etal (1998).
Using definitions for the $J$ and $K$ constants from these papers, if $\eta\ll1$ then
$q^{x} = -K_1 + O(\eta^2 \omega^2)$, $q^r = -J_1 \eta \omega + O(\eta^3 \omega^3)$,
$D^{rr} = {K_1}/{\lambda} + O(\eta^2 \omega^2)$, $D^{rx} = 
-\eta\omega(J_2-J_1K_1) + O(\eta^3 \omega^3)$ and $ D^{xx} =K_2+ O(\eta^2 \omega^2)$.
At the other extreme, for $\eta\gg1$ we have the asymptotic solution 
$q^{x} = O\left({\eta^{-2} \omega^{-2}}\right)$, 
$q^r = -\frac{2}{3}{ \eta^{-1} 
\omega^{-1} }+ O\left({\eta^{-3} \omega^{-3}}\right)$, $D^{rr} = \frac13 + O\left({\eta^{-2} 
\omega^{-2}}\right)$, $D^{rx} =  
O\left({\eta^{-3} \omega^{-3}}\right)$ and $D^{xx} =\frac13 +O\left({\eta^{-2} 
\omega^{-2}}\right)$.

Substituting $\omega = 4r$ and omitting higher orders for clarity obtains, for $\eta \ll 1$,
\be \label{smalleta}
q^r = -4J_1 \eta r,\mbox{~~} q^{x} = -K_1 ,\mbox{~~} D^{rx} =  G_1\eta r,\mbox{~~} D^{rr} = 
{K_1}/{\lambda}, \mbox{~~} D^{xx}=K_2,
\ee
where $G_1=-4(J_2-J_1 K_1) $, and, for $\eta \gg 1$,
\be \label{largeeta}
q^r = -\frac{1}{6}\frac{1}{ \eta r},
\mbox{~~} q^{x} = 0 = D^{rx},\mbox{~~}D^{rr} = \frac13 =D^{xx}.
\ee

\subsubsection{Drift and effective diffusivity, $\eta \ll 1$}

In this limit the cells are less affected by the flow and prone to swim upwards.  Using (\ref{smalleta}) and defining 
$r_0^2= K_1/(2 J_1 \lambda \beta \eta)$,
equations (\ref{lambdastar}), (\ref{R00}) and (\ref{m0star})
provide
\be\label{lambdastarsmalleta1}
\Lambda^*_0(r) = m^*_0(r)\,\left[2 G_1\eta+\Pe(1-2 r_0^2) -K_1\beta \right]+ 2 r^2 R_{0}^0 \left[\Pe\,r_0^2- G_1\eta\right],
\ee
%
%
\be\label{R00smalleta}
R_0^0(r)=\frac{e^{-(r/r_0)^2}}{r_0^2[1-e^{-(1/r_0)^2}]}
%
%
\mbox{~~~~and~~~~}
m_0^*(r)
=\frac{1-e^{-(r/r_0)^2}}{1-e^{-(1/r_0)^2}},
\ee
which satisfies $m_0^*(1)=1$, as required.  
Hence, in the limit $\eta\ll1$ the drift, $\Lambda_0$, is
\be\label{lambdastarsmalleta2}
\Lambda_0 = \Lambda_0^*(1) =
2 G_1\eta\left(1-R_0^0(1)\right)+\Pe\left[1-2 r_0^2\left(1-R_0^0(1)\right) \right] -K_1\beta, 
\ee
highlighting the contributions of swimming diffusion, advection and upswimming.


In a similar manner, the expression (\ref{Deff}) for the effective diffusivity becomes 
\bea
D_{e}&=& - 2 G_1 \eta a_0 + [2\Pe r_0^2 - 2 a_1 G_2] I_1(1) - 2 \Pe I_3(1) + K_2, \label{Deff3}
\eea
where $a_0=(J(1)-\Phi(1))a_1$, $a_1=R_0^0(1)$, $I_i(r)=2\int_0^r {\r}^i (J(\r)-\Phi(\r))R_{0}^0(\r)\,d\r$, for $i=1,3$, and
$G_2=\Pe \,r_0^2-G_1 \eta$. Equation (\ref{Phidef}) yields
\be\label{Jsmalleta}
J(r)=\frac{\lambda}{2 K_1} G_1 \eta\,r^2 \mbox{~~~~and~~~~}
\Phi(r) = \frac{\lambda}{2 K_1}G_2  \left[ r^2 - 2  a_1 \Phi_0(r) \right], 
\ee
where $ 
\Phi_n(r) = \int^r_0 \frac{m_0^*(\r)}{\r R^0_0(\r)^{1-n}} d\r$, for $n=0,1$.
Therefore, for $G_3=G_1 \eta-G_2$,
\be\label{J-Phismalleta1}
(J-\Phi)(r) =  \frac{\lambda}{2 K_1}\left(G_3 r^2 + G_2\,2 a_1\Phi_0(r)\right).
\ee
Some algebra reveals that $I_n(1)=({\lambda}/{2 K_1}) r_0^2 \left[G_3\,I_{n,1}+G_2\,2 a_1\,I_{n,2}\right]$, where 
$I_{1,1}=1- a_1$, 
$I_{1,2}=\Phi_1(1)-a_2$, 
$I_{3,1}=2r_0^2I_{1,1}- a_1$, 
$I_{3,2}=r_0^2\left[I_{1,2}-\frac{1}{2}I_{1,1}\right]+\frac{1}{2}-a_2$,
and $a_2=a_1\Phi_0(1)$.
Hence,
\be\label{Desmalletasmallr0-3}
D_e = K_2 + \frac{\Pe}{\beta} \frac{1}{2 J_1}\left( G_1 b_2(r_0^2) +  \frac{\Pe}{\beta} \frac{1}{2 J_1} \frac{K_1}{\lambda} \frac{1}{\eta^2} b_3(r_0^2)\right)+ \frac{\lambda}{K_1}\,G_1^2 \eta^2  b_1(r_0^2),
\ee
where, recalling that $r_0^2=K_1/(2 J_1 \lambda \beta \eta)$,
\bea
b_1(r_0^2)&=&2 a_1(a_2-1)+r_0^2 2 a_1(I_{1,1}-a_1 I_{1,2})],\nnn
b_2(r_0^2)&=& a_1(1-2a_2+2I_{3,2})-2 I_{3,1}
+r_0^2\{a_1[2 I_{1,2} (2 a_1-1) -3 I_{1,1}]+2 I_{1,1}\}, 
\,\,\,\,\,\, \nnn
b_3(r_0^2)&=& I_{3,1}-2a_1 I_{3,2} + r_0^2 \{a_1[2 I_{1,2} (1-a_1) +I_{1,1}]- I_{1,1}\}.
\eea


\subsubsection{Drift and effective diffusivity, $\eta\gg1$}

In this limit, the cells are affected by the flow to the extent that they mostly tumble.
Using (\ref{largeeta}), equations (\ref{R00}), (\ref{m0star}) and (\ref{lambdastar})
provides $R_0^0(r)=\left(1-\nuu\right) r^{-2 \nuu}$,
where $\nuu=\beta/4\eta$, $m_0^*(r)=r^{2(1-\nuu)}$ and 
\be\label{lambdastarlargeeta2}
\Lambda^*_0(r) =  \Pe\,r^{-2 \nuu}\left(r^2-\frac{2(1-\nuu)}{2-\nuu}r^4\right),
\ee
respectively.  Hence,
\be\label{driftlargeeta2}
\Lambda_0 =\Lambda^*_0(1) = \Pe \frac{\nuu}{2-\nuu}.
\ee


Similarly, with definitions (\ref{largeeta}), equation (\ref{Phidef}) yields
\be\label{Philargeeta1}
\Phi(r) =  \Pe\,\frac{3}{2(2-\nuu)}\left(r^2-\frac{r^4}{2}\right).
\ee
Hence, the equation (\ref{Deff}) for the effective diffusivity gives
\be
D_e =   \frac{1}{3} + 2(\Lambda_0 - \Pe) \int_0^1 r \Phi R_0^0\,dr + 4\Pe \int_0^1 r^3\Phi 
R_0^0\,dr = \frac{1}{3} + \Pe^2 G(\nuu), \label{Delargeeta1}
\ee
where 
\be
G(\nuu)=\frac{3}{2}\frac{1-\nuu}{2-\nuu}\left[\frac{1-\nuu}{2-\nuu}\left(\frac{1}{3-\nuu} 
-\frac{2}{2-\nuu}\right)+\frac{2}{3-\nuu}-\frac{1}{4-\nuu} \right].
\ee

\subsubsection{Dependence of dispersion on flow parameters in the strong and weak limits}

\begin{figure}
\begin{center}
a)\rotatebox{0}{\includegraphics[width=5.7cm]{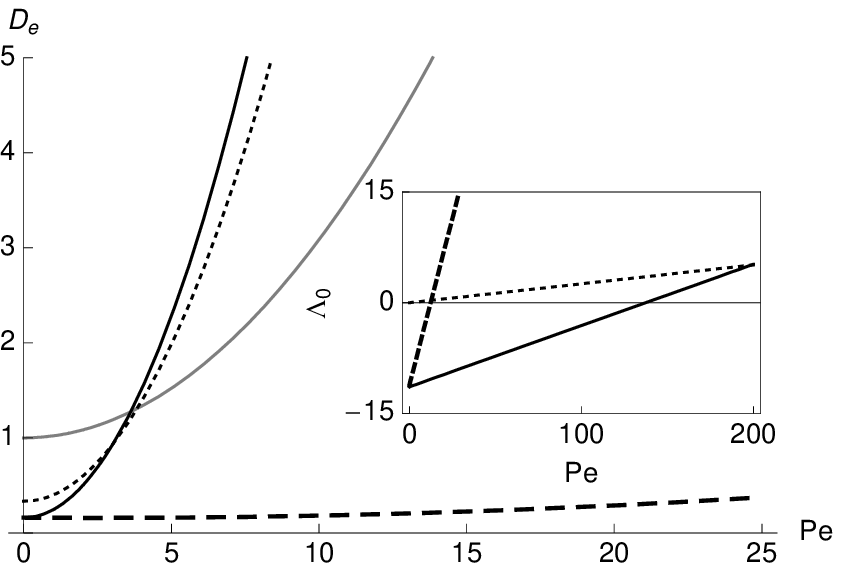}}~~
b)\rotatebox{0}{\includegraphics[width=5.7cm]{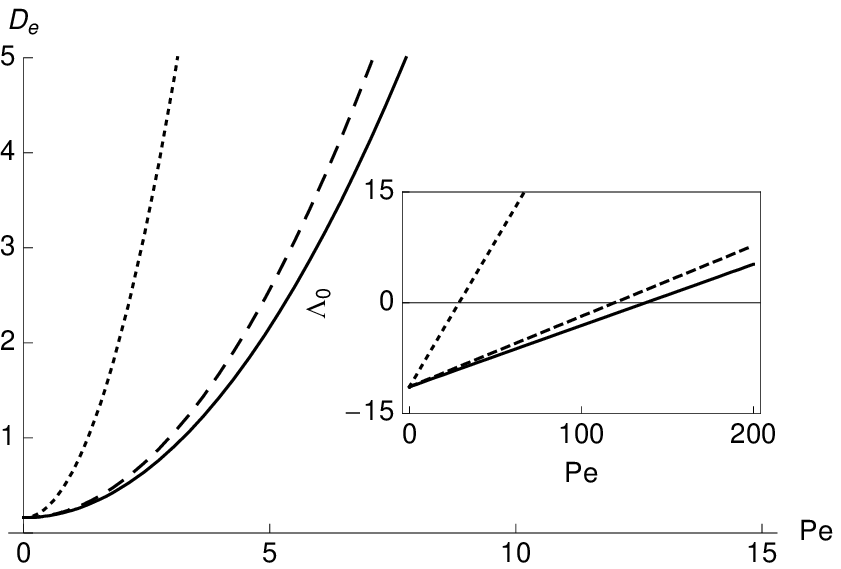}}\\
\caption{{\bf Effective diffusivity (inset: drift, $\Lambda_0$) against $\Pe$
calculated using asymptotic solutions to the Fokker-Planck approach. a) Poiseuille
approximation (case I) for $\eta=0.007$ (solid line), $\eta=0.3$ (long
dashed), $\eta=100$ (dashed) and the classical Taylor-Aris result
(grey). b) Self-driven flow ($\tilde{p}_x=0$) using the coupled
solution from case II (broad plumes; $\eta=0.007$). Mode 1 (long dashed) and mode 2 (dashed)
are shown with the uncoupled limit (solid).} \label{f:effDplots}}
\end{center}
\end{figure}

Here, drift and diffusivity are evaluated as a
function of Pe for realistic parameters.
Recalling $\Pe=U a/D^c$, $\beta=V_s a/D^c$, $\eta=U B/a$ and
$\lambda=1/(2Bd_r)$ (where $d_r$ is the rotational diffusion
constant for swimming cells) we see it is in theory possible
to vary $\Pe$ whilst holding $\beta$, $\eta$  and $\lambda$
(and so $r_0,\,\nuu$) fixed. 
For \CN\, the gyrotactic reorientation time $B=3.4$ s,  $d_r=0.067$ s$^{-1}$, and so
$\lambda=2.2$, thus $K_1=0.57$, $K_2=0.16$, $J_1=0.45$,
$J_2=0.16$ (Pedley \& Kessler 1990; Hill \& H\"{a}der, 1997). 
With these values,
$G_1=-4(J_2-J_1K_1)=0.39$. Furthermore, the average swimming
speed and cell diffusivity are
$V_s\approx$ $10^{-2}$ cm s$^{-1}$ and
$D^c\approx5\times10^{-4}$ cm$^2$ s$^{-1}$ (Hill \& H\"{a}der
1997, Vladimirov \etal 2004). Using these parameters and
$a = 1$ cm, we find that $\beta=20$ and $r_0=\sqrt{2}$ for
$\eta=0.007$, $r_0=0.22$ for $\eta=0.3$, and
$\nuu=\beta/(4 \eta)=0.05$ for $\eta=100$.  Hence, expressions (\ref{lambdastarsmalleta2}), (\ref{Desmalletasmallr0-3}),
(\ref{driftlargeeta2}) and (\ref{Delargeeta1}) are used to plot 
the effective diffusivity and drift (inset) for algae in a
Poiseuille flow in figure \ref{f:effDplots}a.
The figure reveals that low and high levels of gyrotaxis (measured by $\eta$) lead to behaviour 
akin to Taylor dispersion but intermediate levels dramatically reduce the impact
of advection.  This is because at these intermediate levels the cells form dense plumes
in the centre of the tube so are not subject to the full range of flow speeds.
On the other hand, intermediate gyrotaxis does lead to large amounts of 
swimming and flow induced drift relative to the mean flow, due to their central location.  
For small and intermediate $\eta$ the asymptotic results reveal that the drift changes sign for a non-zero Pe number.  However, the asymptotic results for large $\eta$, are not strictly valid for small Pe.  Nonetheless, one would expect the drift to change sign in a similar manner, such that all three curves intersect on the $y$-axis.

\subsection{Algae in self-driven flow (weak coupling, $A\ll1$)}

Recall from section \ref{s:flow} that for self-driven
flows there are two solutions: a simple
mode 1 flow
and a mode 2 flow with upwelling at
the tube sides.
To calculate the transport
coefficients in these cases the definitions
(\ref{smalleta}) are employed (since the flow solutions were all
obtained for $\eta\ll1$ and weak coupling, $A\ll1$).
With the same parameters as for the  $\eta=0.007$ case in figure \ref{f:effDplots}a, 
figure \ref{f:effDplots}b plots the diffusivity and drift (inset).
It is clear that the mode 1 results for these broad plumes are rather similar to those generated by Poiseuille flow.  However, for the mode 2 solutions cells both drift and diffuse faster, likely due to the greater shear.

\section{Discussion} \label{s:discussion}

In this paper, we derive exact expressions in the long-time
limit for the mean drift and effective axial diffusion of an
axisymmetric blob of biased, swimming microorganisms in a
plume in a pipe flow driven by an external pressure gradient
and the presence of the (negatively) buoyant cells. In the
same limit, we find that the axial skewness of the
cross-sectionally averaged cell distribution vanishes. The
results are independent of the cell geometry, swimming
behaviour and model used to represent the cell-flow
interactions.

Explicit results for several useful cases are presented,
from the Taylor-Aris limit to fully coupled gyrotactic
spherical swimming cells (i.e.~cells that drive the flow and
whose swimming direction is biased by external and viscous
torques).  The expressions reveal the mechanisms for several
competing effects and explain how these lead to diffusion
and (positive or negative) drift through the tube.
Fundamentally, the cells swim and, in the limit that they
are very bottom heavy, they may swim mostly against a
downwelling flow, leading to a negative drift relative to the
mean flow.  On the other hand, cells that are not
bottom heavy act more like diffusing passive tracers, with no drift.  In
both these cases the cells diffuse as for Taylor-Aris
dispersion. However, an intermediate degree of bottom
heaviness leads to much more interesting behaviour. A balance
between gravitational and viscous torques, a balance that will vary
across the pipe flow, can lead the cells to form gyrotactic
plumes, inducing further flow and self-concentration.  These centrally focused
plumes of cells can be strongly advected with the flow
(i.e.~faster than the mean flow) but will sidestep classical
shear-induced Taylor-Aris dispersion; effective diffusion may be dominated by swimming
diffusion even for large flow rates. It is clear that
swimming behaviour leading to drift across streamlines can 
have a tremendous influence on cell transport in such systems.

The results are sufficiently general that they may easily be
applied to other micro-organisms and taxes, such as
chemotaxis in suspensions of bacteria swimming in flows in
microfluidic chambers, or spermatozoa {\it in vivo}.  In a
subsequent paper we shall provide further explicit examples
for non-spherical cells (behaviour influenced by the
rate-of-strain tensor) and  for additional swimming stresses
for concentrated suspensions.  Both these aspects will
modify the plume structure and thus affect axial cell
transport.

Work in progress is exploring how the theory can be applied
to determine the qualitative form of the orientationally
averaged cell swimming diffusion tensor for suspensions of
gyrotactic cells from experiments. For a realizable
experiment one must introduce dyed cells into a plume whilst
maintaining a constant cross-sectionally averaged cell
concentration.  This may be achieved simply by momentarily
switching from undyed to dyed cells at the input or using
photoactivatable GFP for localized photolabelling of cells
(Patterson \& Lippincott-Schwartz 2002). Note that plume
solutions for the various diffusion descriptions differ
qualitatively for large Peclet numbers, and thus so must
predictions for mean drift and effective diffusion.  Hence,
we aim to clarify the applicability of differing diffusion
approximations in a general shear flow.


\section{Acknowledgements}

The authors gratefully acknowledge support from the EPSRC (EP/D073398/1).

\end{document}